# Les Nouvelles Voies de Financement Novatrices : Vecteur de Changement Transformateur vers la Performance Financière des Entreprises au Maroc

# Innovative Financing Solutions: A Transformative Driver for Financial Performance of Businesses in Morocco

**BADRANE Nohayla**
Doctorante
Ecole Nationale de Commerce et de Gestion
Université Hassan Premier Settat, Maroc
Laboratoire de recherche en finance, audit et gouvernance des organisations

**BAMOUSSE Zineb**
Enseignante chercheuse
École Nationale de Commerce et de Gestion
Université Hassan Premier Settat, Maroc
Laboratoire de recherche en finance, audit et gouvernance des organisations

**Résumé**

Sous l'emprise d'un paradigme économique traversé par des mutations constantes et des défis contemporains complexes, la gestion de la trésorerie s'impose la pierre angulaire qui conditionne la pérennité et influe sur la performance des entreprises. Face à ces enjeux, les trésoriers sont appelés à se servir des modes de financement innovants tels que le capital-risque, la finance verte, le Crowdfunding, les solutions avant-gardistes des banques panafricaines, ainsi que les services offerts par la blockchain. Ces outils novateurs agissent comme des catalyseurs majeurs de la résilience face à un paysage imprégné de la volatilité des marchés, des transitions écologiques et des exigences croissantes d'adaptation technologique. A ce propos, le présent article a pour ambition d'investiguer la manière dont ces modes de financements innovants peuvent servir de leviers stratégiques pour transformer ces défis en opportunités. Les résultats mettent en évidence que l'innovation au cœur de la gestion de la trésorerie constitue un vecteur incontestable pour soutenir la compétitivité et booster la performance des entreprises marocaines, Elle ouvre ainsi de nouvelles perspectives de développement en leur permettant de naviguer avec agilité à travers les turbulences d'un environnement en pleine mutation.

**Mots-clés :** Gestion de trésorerie innovante ; Performance financière ; Pérennité ; Instruments de financement innovants ; résilience.

**Abstract**

In a rapidly evolving landscape marked by continuous change and complex challenges, effective cash management stands as a cornerstone for ensuring business sustainability and driving performance. To address these pressing demands, cash managers are increasingly turning to innovative financing solutions such as venture capital, green finance, crowdfunding, advanced services from Pan-African banks, and blockchain technology. These cutting-edge tools are pivotal in bolstering resilience against market volatility, ecological transitions, and the accelerating pace of technological change. The present article aims to examine how such innovative financial approaches can serve as strategic drivers, enabling businesses to transform challenges into opportunities. The analysis underscores that rethinking cash management through innovation is a critical pathway to boost the performance of Moroccan companies. Therefore, embracing these forward-thinking strategies unlocks new avenues for development empowering them to adapt with agility amidst the uncertainties of a shifting environment.

**Introduction**

La performance et partant la survie des entreprises suscitent l'attention de l'échiquier international, dans la mesure où elles sont universellement perçues comme les piliers de l'essor économique et du développement social. Cette vérité universelle trouve un écho particulier au Maroc, où la quête de la performance se dresse comme un horizon désiré, en particulier dans un contexte marqué par des mutations incessantes.

Au sein de ce panorama, les entreprises marocaines, au centre névralgique de l'économie nationale, recèlent un potentiel important de croissance à mettre en valeur. À la croisée de cette dynamique repose la gestion de la trésorerie, pierre angulaire qui conditionne leur pérennité et influe sur leur performance. Toutefois, bien que ces organisations représentent l'épine dorsale de l'économie marocaine, de divers défis se dressent devant leur évolution. Ces difficultés, qui pèsent sur la performance et grèvent lourdement la trésorerie, rendent impérative une réflexion profonde sur les stratégies de gestion de la trésorerie. Dans ce contexte, l'innovation au cœur des modes de financement apparaît comme une clé de voûte pour naviguer à travers les eaux tumultueuses d'un environnement globalisé. A cet effet, nous soutenons l'idée selon laquelle la performance financière des entreprises est tributaire d'une gestion de trésorerie innovante qui intègre dans ses pratiques des options de financement avant-gardistes.

Sous cet angle, les conclusions issues de cette étude révèlent qu'un mode de financement innovant constitue un gage de pérennité contre les risques de défaillance et s'impose comme un véritable pivot de la performance des entreprises marocaines, paramètre important de leur équilibre financier et vecteur catalyseur de leur croissance.

L'ambition de cette étude se déploie en deux axes majeurs. Premièrement, elle entend souligner la portée stratégique du financement dans la croissance et la survie des entreprises en explorant les avenues de financement innovantes à leur disposition. Deuxièmement, elle aspire à mettre en exergue la nécessité d'intégrer l'innovation au cœur des stratégies de la fonction de la trésorerie, spécifiquement dans le choix des financements, en soulignant les apports prometteurs sur la performance des entreprises au Maroc.

Sous ce prisme, le présent article s'efforce de fournir des pistes de réflexion à l'interrogation suivante : En quoi les instruments de financement innovants peuvent-ils constituer un vecteur de changement transformateur vers la performance des entreprises au Maroc ?

La question explorée dans cette étude pousse à avancer l'hypothèse fondamentale suivante :

H1 : Mobiliser les modes de financement innovants en faveur de la gestion de trésorerie contribue à booster la performance financière des entreprises marocaines.

De cette hypothèse centrale émergent six sous-hypothèses, détaillées ci-après :

H1-1 les alternatives de financement novatrices permettent aux trésoriers de contourner les difficultés de financement.

H1-2 la palette des financements innovants s'annonce un vecteur incontournable pour alléger les coûts cachés.

H1-3 Un mode de financement innovant assure une gestion optimisée des coûts.

H1-4 les modalités de financement novatrices s'affirment un vecteur catalyseur de la croissance.

H1-5 Un choix de financement innovant rehausse la réputation de l'entreprise en tant qu'un partenaire clé dans les causes environnementales et sociales.

H1-6 les instruments de financement novateurs contribuent significativement à garantir la pérennité et renforcer la résilience des entreprises face aux défis contemporains.

Guidé par les hypothèses et la question de recherche précédemment exposées, le présent article est structuré en cinq sections. La première est dédiée à une revue de littérature approfondissant les fondements conceptuels portant sur la gestion de la trésorerie, l'innovation et la performance financière. La seconde présente la démarche méthodologique empruntée. La troisième section mobilise le pouvoir explicatif de la théorie des coûts de performances cachées pour élucider le sujet abordé. La quatrième section dévoile les résultats obtenus et la discussion qui en découle. La dernière section, quant à elle, synthétise les apports managériaux, académiques et socio-économiques en identifiant les limites et les pistes de réflexion pour des recherches futures.

## 1. Revue de Littérature et Fondements Conceptuels de la Gestion de la Trésorerie, la Performance Financière et l'innovation

Cette première partie vise à fournir un éclairage tout en établissant le lien entre les concepts clés à savoir la gestion de la trésorerie, l'innovation et la performance financière qui sous-tendront les développements ultérieurs.

### 1.1. La Gestion de la Trésorerie

La gestion de la trésorerie est un concept multidimensionnel, se manifestant à travers différentes pratiques, mais qui converge vers une compréhension commune de son rôle fondamental au sein de la finance d'entreprise. Ainsi, Bruselerie et Eliez (2017) mettent en avant que la gestion de la trésorerie soit la pierre angulaire de la fonction financière de l'entreprise. Elle en constitue le bras armé. Autrefois, reléguée à un simple élément de la direction financière de l'entreprise, elle est désormais reconnue comme une entité cohérente et multiforme endossant un rôle stratégique croissant. En outre, Desbrières et Poincelot (2015) soulignent que la gestion de la trésorerie englobe l'ensemble des décisions, règles et procédures permettant d'assurer au moindre coût le maintien de l'équilibre financier instantané de l'entreprise. Sous cet angle, le rôle du trésorier se cristallise autour de deux missions principales à savoir : la gestion des risques financiers et la gestion des liquidités en veillant à ce que l'entreprise dispose d'une encaisse suffisante pour honorer, au moindre coût, ses obligations financières prévues. Derrière cette définition lapidaire, se profile l'accomplissement de deux responsabilités fondamentales. La première consiste à assurer les prévisions de trésorerie fiables indépendamment de l'horizon temporel envisagé. Ensuite, la seconde mission consiste à sélectionner de façon optimale les décisions de financements. Ainsi l'article en question met en avant l'importance de choisir des modes de financements innovants en vue de propulser la performance financière des entreprises.

Dans la même veine, Ross et ses co-auteurs élargissent la portée de la gestion de la trésorerie en l'inscrivant dans une perspective plus stratégique. Pour ces auteurs, la gestion de la trésorerie dépasse la simple gestion des flux de liquidités et s'étend à l'optimisation des ressources financières pour maximiser le rendement tout en atténuant les risques inhérents aux fluctuations du marché (Ross, et al., 2019).

Sous le même angle d'optimisation, (Reyes et Soria 2021), mettent en relief que l'objectif stratégique de la gestion de la trésorerie s'appuie sur la réduction des fonds dédiés au financement externe, la diminution des capitaux immobilisés et la compression des coûts liés aux transactions monétaires. A cet égard, une gestion de trésorerie optimisée est un préalable incontournable à la pérennité de toute organisation. Dès lors, l'incapacité à mettre en œuvre des pratiques de gestion de trésorerie adaptées peut gravement compromettre la performance de l'entreprise (Kithinji, 2023).

Sur la base des réflexions précédemment énoncées, le rôle crucial de la trésorerie dans la survie de toute entreprise ne saurait être contesté. Cette importance trouve un écho dans les propos de Ludwig Von Bertalanffy. Selon lui, la trésorerie est à l'entreprise ce que le sang est à l'organisme : sans elle, l'entreprise ne peut survivre. En conséquence, la trésorerie n'est plus perçue comme un simple support opérationnel, mais comme un ancrage indispensable à la performance durable. Elle se positionne ainsi comme un acteur central dans l'architecture financière, et sa valeur se confirme davantage à mesure que les entreprises cherchent à anticiper les risques et à saisir les opportunités de croissance.

Dans cette dynamique de transformation, (KPMG, 2024) met en lumière l'ampleur stratégique croissante de la gestion de trésorerie, un levier désormais indissociable de la performance et de la résilience des entreprises. A ce titre, l'étude de KPMG approfondit la manière dont une gestion proactive de la trésorerie permet aux entreprises de naviguer dans l'incertitude, en transformant les risques en opportunités de croissance. En effet, une gestion de trésorerie bien orchestrée permet aux entreprises de se frayer un chemin à travers les incertitudes économiques, de saisir de nouvelles opportunités et de propulser la rentabilité tout en mitigeant les risques financiers.

Il est dès lors essentiel d'examiner la gestion de la trésorerie sous le prisme des mutations accélérées de l'environnement financier vers la globalisation, en démontrant qu'une gestion de trésorerie rigoureuse et avant-gardiste contribue significativement à booster la performance notamment financière des firmes marocaines. Cette performance financière peut être impulsée à travers un choix de financement innovant susceptible de maintenir l'équilibre financier tout en assurant la pérennité des entreprises dans l'arène nationale.

### 1.2. La Performance Financière

Avant de présenter le cadre théorique relatif à la performance financière, il importe de signaler que le consensus semble impossible pour forger une définition unanime autour de la performance, ce qui révèle le caractère polysémique de cette notion lourdement investi par plusieurs disciplines notamment par la gestion financière. Dans ce contexte, la performance est toute action orientée vers la concrétisation des ambitions stratégiques. A ce propos, la performance est tout ce qui concourt à améliorer le couple valeur-coût (Lorino, 2003). Par ailleurs, la performance est un concept multidimensionnel qui recouvre les notions d'efficacité et d'efficience. Elle est polysémique dans le sens où elle englobe une diversité de dimensions telles que la dimension commerciale, économique, sociale, sociétale, environnementale…Cet article se concentre sur la dimension financière vu qu'elle est intimement liée aux choix de financement et s'impose comme une dimension fédératrice des autres. A cet égard Bouquin dans son ouvrage intitulé « le contrôle de gestion » privilégie la perspective financière et associe la performance à trois piliers clés à savoir : l'économie, l'efficience et l'efficacité. L'économie repose sur l'acquisition de ressources au moindre coût. L'efficience, quant à elle, établit une corrélation entre les résultats obtenus et les moyens mobilisés. L'efficacité à son tour reflète la capacité de l'entreprise à atteindre les objectifs et les finalités poursuivis (Bouquin, 2004).

Dans le même ordre d'idées, la performance financière illustre l'efficacité dans la mobilisation et l'emploi des ressources financières (Martory et Crozet, 2016). Cette approche met en exergue l'importance cruciale d'une gestion optimale des capitaux pour l'atteinte des objectifs stratégiques d'une entreprise. Par ailleurs, la performance financière est associée à la réalisation d'une rentabilité soutenue, d'une croissance satisfaisante et d'une création de valeur pour l'actionnaire (Guérard, 2006). En outre, il convient de souligner que le dégagement des résultats défavorables ou insuffisants traduit un gaspillage de ressources dommageables pour l'entreprise. De ce fait, le maintien des performances financières à un niveau satisfaisant est une mission fondamentale assignée à la gestion financière (Cohen, 1991).

Dans la même lignée de réflexion, la performance financière tend à stabiliser le financement en réduisant le recours aux crédits. Cet objectif est un impératif majeur assigné aux trésoriers qui sont appelés à optimiser la gestion de la trésorerie en intégrant des pratiques novatrices en termes de financement afin d'accroitre la performance financière de leurs firmes.

### 1.3. L'innovation

L'innovation, telle que définie par le manuel Oslo relevant de l'organisation de coopération et de développement économique (OCDE) est : « la mise en œuvre d'un produit, d'un processus nouveau ou sensiblement amélioré, d'une nouvelle méthode de commercialisation ou d'une nouvelle méthode organisationnelle dans les pratiques de l'entreprise, l'organisation du lieu de travail ou les relations extérieures. » Cette définition d'ensemble couvre un large éventail d'innovations à savoir les innovations de produit, de procédé, de commercialisation et d'organisation (OCDE, 2005 : 54).

Le présent article met un accent particulier sur l'innovation managériale, étant donné que les choix optimaux et innovants en matière de financement exigent, sans équivoque, une capacité d'innovation managériale significative. Cette capacité à innover dans la gestion de la trésorerie devient le bouclier indispensable pour impulser la performance financière et catalyser la résilience de l'entreprise.

A cet égard, Mol et Birkinshaw (2009) associent l'innovation managériale à la performance en postulant que l'innovation managériale est la mise en œuvre des pratiques de management nouvelles pour l'entreprise dans l'intention d'accroitre sa performance. Dans le même ordre d'idées, Damanpour et Aravind établissent un lien direct entre l'innovation managériale et la performance en considérant que « l'innovation managériale est une nouvelle organisation, un nouveau système administratif, de nouvelles pratiques managériales, ou de nouvelles techniques susceptibles de créer de la valeur pour l'organisation qui les adopte » (Damanpour et Aravind, 2012 : 424). Cette définition met en lumière l'essence cruciale de l'innovation, qui transcende la simple sphère des opérations ou des produits pour s'ancrer profondément dans le cœur même de la gestion d'une entreprise et ses stratégies déployées y compris celles liées au financement.

Dans le même sillage, Jaouen et Le Roy estiment que l'innovation managériale touche l'organisation dans son ensemble dans le sens où toutes les fonctions de l'entreprise sont constamment remodelées par les innovations managériales (Jaouen et Le Roy, 2013).

Il ressort de ce qui précède que toutes les fonctions de l'entreprise doivent s'immerger dans le courant de l'innovation. Dans ce contexte, la fonction financière à sa tête la gestion de la trésorerie n'est pas épargnée, mais elle figure dans le haut du palmarès des fonctions les plus concernées par cette transformation. En effet, si l'innovation a longtemps été la chasse gardée des départements Recherche et Développement, l'heure est venue de l'étendre pour pénétrer la fonction de la trésorerie en vue d'en faire un atout catalyseur de la performance. Ainsi, dans un monde volatil et globalisé, les entreprises sont invitées à embrasser le rivage numérique, en instaurant une culture où prévaut l'innovation, une culture qui crée un héritage durable pour l'organisation. Parallèlement, les trésoriers, au cœur de cette mutation, sont invités à repenser leur ADN de gestion vers des pratiques plus novatrices afin de renforcer leur résilience face à cette multitude de défis pouvant compromettre leur pérennité (Badrane et Bamousse, 2023).

## 2. Objectif et Méthodologie

Dans le cadre de ladite étude, nous nous inscrivons dans un domaine où la gestion de la trésorerie éveille un intérêt significatif dans la littérature financière. Cependant, bien que nombreux travaux aient exploré ces thématiques sous différents angles, force est de constater que l'innovation dans la gestion de la trésorerie, en particulier dans les modes de financement, reste un champ relativement peu couvert. Cette lacune servira de socle à l'élaboration de cet article qui vise à forger un pont entre les pratiques innovantes de la gestion de la trésorerie et l'amélioration de la performance financière des entreprises.

Pour y parvenir, une approche analytique rigoureuse a été adoptée, reposant sur un examen approfondi de la littérature scientifique, des ouvrages de référence dans le domaine de la finance d'entreprise et des travaux de recherche issus de sources académiques renommées. Les articles, soigneusement sélectionnés de bases de données réputées telles que Cairn.info, Springer, Jstor, Elsevier et EBSCO…ont été complétés par des rapports et des études provenant d'institutions internationales, reconnues notamment l'OCDE, Nations Unis commission économique pour l'Afrique, PWC France, KPMG, la Banque Mondiale … Cette démarche holistique a permis de disposer d'un large spectre d'informations couvrant à la fois les aspects théoriques et pratiques des thématiques abordées.

L'analyse a pris appui sur quatre volets essentiels : la gestion de la trésorerie, l'innovation, la performance financière des entreprises, et l'apport des modes de financement innovants.

L'objectif est de démêler les fils reliant les stratégies innovantes de gestion de trésorerie, en particulier celles liées choix de financement, à une amélioration tangible de la performance des entreprises. Parallèlement, ladite étude s'est concentrée sur la manière dont ces pratiques, en se détachant des méthodes conventionnelles, peuvent servir de levier pour dynamiser la croissance, maintenir l'équilibre financier, renforcer la résilience et, in fine, garantir la pérennité des entreprises marocaines sur le podium économique national.

Parallèlement, en vue de souligner avec précision l'importance des modes de financement innovants dans l'allègement du fardeau financier global, nous avons spécifiquement mobilisé le pouvoir explicatif de la théorie des coûts de performances cachées. Cette théorie, en se concentrant sur les coûts souvent omis ou sous-estimés dans les évaluations conventionnelles de la performance, fournit un cadre explicatif puissant pour examiner comment un mode de financement innovant peut contribuer à la réduction de ces coûts cachés associés au financement. Ceci contribue de manière significative à améliorer l'efficacité opérationnelle et, par extension, à stimuler la performance des entreprises.

La méthodologie empruntée va au-delà d'une simple revue de la littérature existante ; elle aspire également à contribuer de manière significative au débat académique en établissant un dialogue entre les travaux précurseurs et les dynamiques actuelles du monde des affaires. Par ce biais, l'étude entend forger des recommandations pratiques et esquisser de nouvelles orientations pour les investigations ultérieures, pavant ainsi la voie vers une gestion de trésorerie plus dynamique et adaptée aux défis contemporains.

### 3. Motifs Théoriques

Pour décrypter l'impact d'un mode de financement novateur sur l'allègement des coûts et, par voie de conséquence, sur l'accroissement de la performance des entreprises marocaines, la présente étude s'adosse à la théorie des coûts de performances cachés, élaborée par Henri Savall. Cette approche théorique, en mettant l'accent sur les coûts souvent invisibles mais profondément impactant la trésorerie des entreprises, offre un cadre d'analyse enrichi pour comprendre comment les décisions de financement peuvent transformer ces coûts latents en bénéfices tangibles.

Selon Savall, les coûts cachés se manifestent à travers divers canaux au sein de l'entreprise, englobant les inefficacités opérationnelles, les délais, les erreurs ou encore les opportunités manquées, qui, ensemble, représentent un fardeau significatif sur la performance globale (Cappelleti, et al., 2018). Dans le contexte de la présente étude, mobiliser la théorie des coûts de performances cachées permet d'éclairer spécifiquement la manière dont les choix de financement innovants peuvent atténuer ces coûts. En se détournant des voies traditionnelles de financement au profit d'options plus novatrices telles que le Crowdfunding, le capital risque ou encore la finance verte, les entreprises marocaines se dotent d'une flexibilité inédite. Ce choix leur permet non seulement de contourner les écueils liés aux coûts d'opportunité et de transaction, mais aussi de valoriser leur image sur l'échiquier national. Cet aspect réputationnel est essentiel dans un marché de plus en plus attentif aux pratiques éthiques et durables.

Par cette réorientation stratégique, les entreprises bénéficient de conditions plus flexibles réduisant ainsi significativement les coûts d'opportunité. Cela leur permet de réallouer des ressources précieuses vers des initiatives stratégiques, plutôt que de les immobiliser dans des emprunts onéreux. Parallèlement, le déploiement de technologies de pointe notamment les plateformes numériques et la blockchain dans les processus de financement optimise l'efficacité opérationnelle. Cette modernisation des processus réduit les coûts de transaction, minimisant ainsi les frais administratifs et les délais souvent associés aux modalités de financement traditionnelles. Au-delà de ces avantages directs, l'engagement dans des modes de financement à la fois innovants et socialement responsables, à l'instar de la finance verte, confère aux entreprises une meilleure réputation. Cette perception positive peut attirer des conditions de financement plus favorables à long terme, réduisant les coûts cachés liés à la recherche et à la négociation de financement.

En somme, l'adoption de modes de financement avant gardistes représente une stratégie cruciale pour les trésoriers cherchant à garantir la disponibilité des fonds au moment opportun. Contrairement aux méthodes de financement traditionnelles, qui peuvent être entravées par des processus d'approbation longs et des conditions rigides, ces alternatives offrent une souplesse essentielle pour répondre rapidement aux besoins financiers. Cette réactivité permet également de saisir des opportunités de croissance immédiates tout en évitant les manques à gagner potentiellement occasionnés par l'indisponibilité de liquidités.

Ainsi, en mobilisant la théorie des coûts de performances cachées, le présent article se propose de démontrer comment une gestion de trésorerie éclairée est capitale pour débloquer les leviers de performance et d'efficacité au sein des entreprises marocaines, offrant un nouveau prisme à travers lequel un mode de financement innovant peut transformer les contraintes financières en véritables opportunités de croissance.

## 4. Résultats et Discussion

### 4.1. L'Apport Stratégique de l'Innovation à l'Optimisation de la Performance Financière des Entreprises

L'innovation avec toutes ses formes constitue un levier de croissance, un pionnier de performance et un catalyseur de développement notamment au regard des transformations accélérées de l'environnement vers la digitalisation. Dans cette perspective, (Gaynor, 2002) met en exergue le rôle de l'innovation dans la pérennité des organisations. Il a souligné que l'intégration de l'innovation dans la culture organisationnelle est fondamentale pour permettre aux entreprises de naviguer à travers les eaux turbulentes du marché contemporain. Ainsi, sans une approche proactive en matière d'innovation, les entreprises risquent de se retrouver dépassées, incapables de s'ajuster aux mutations effrénées de leur environnement. De même, Il soutient que l'innovation n'est pas simplement une initiative ponctuelle, mais un socle incontournable de la culture d'entreprise qui doit être intégrée dans toutes les facettes de l'organisation.

Dans ce contexte, la fonction financière n'est pas épargnée. Elle est dans l'obligation d'innover en se dotant d'un processus souple, d'une structure flexible et d'un système agile en vue de s'adapter aux nouvelles transformations (Barzi et Bamousse, 2023).

Dans cette même veine, l'innovation se révèle une nécessité impérieuse dans les périodes de crise, où les défis deviennent des catalyseurs de changement et de transformation pour les entreprises. C'est dans ce contexte que l'étude d'Adam et Alarifi a pris tout son sens, en démontrant avec acuité comment, face à l'adversité imposée par la pandémie de COVID-19, les PME saoudiennes ont su tirer parti de l'innovation pour non seulement survivre mais aussi prospérer. En se penchant sur les pratiques innovantes adoptées par ces entreprises, l'étude met en avant le rôle primordial de l'innovation dans l'amélioration de la performance et l'augmentation des chances de pérennité en temps de crise. L'adoption de technologies de pointe, L'intégration de nouvelles méthodes de gestion, et l'engagement actif des employés dans le renouvellement des activités de l'entreprise ont constitué le socle sur lequel les PME ont pu s'appuyer pour naviguer à travers la tempête du COVID-19 (Adam et Alarifi, 2021).

Ces initiatives, loin d'être de simples mesures réactives, se sont révélées être des stratégies proactives essentielles, marquant le passage d'une gestion traditionnelle à une posture résolument tournée vers l'avenir, où l'innovation devient le maître-mot de la résilience et du succès durable.

Dans le même sillage, les conclusions des études réalisées dans le contexte tunisien par Chouaibi et ses co-auteurs sur le niveau d'innovation des entreprises et son impact sur la performance financière ont démontré que les activités d'innovation impactent positivement et significativement la performance financière des entreprises. A ce propos, les entreprises les plus novatrices sont mieux positionnées pour générer une rentabilité économique et financière accrue. Cette situation favorable facilite, par conséquent, l'accès à des financements internes et externes à des conditions plus avantageuses (Chouaibi, et al., 2010). Ainsi l'innovation sous cet angle permet non seulement de booster la performance financière mais également de sécuriser la situation financière tout en optimisant les coûts de financement. Et comme aucune entreprise n'est épargnée par le dépôt de bilan, la fonction de la trésorerie doit donc faire preuve d'une grande agilité pour s'adapter aux aléas d'un environnement en constante évolution (Badrane et Bamousse, 2023). Dès lors, il devient impératif pour les trésoriers de repenser leur approche de gestion, en capitalisant sur le potentiel des financements innovants.

Dans le même ordre d'idées, Djoutsa et ses co-auteurs ont mené une étude quantitative sur la base d'un échantillon de 231 grandes entreprises camerounaises ayant investi dans les activités d'innovation durant la période allant de 2004 à 2011. Les résultats obtenus ont révélé que les firmes les plus performantes se distinguent par une forte capacité d'innovation. Ce qui atteste que l'investissement dans les activités innovantes permet à l'entreprise d'améliorer sa position concurrentielle et donc de propulser sa performance financière (Djoutsa, et al., 2017).

Il ressort de ce qui précède que la littérature existante a indéniablement souligné le rôle déterminant de l'innovation, considérée dans un sens élargi, comme un levier majeur pour la survie et le développement des entreprises. Cette reconnaissance traverse divers domaines, allant de la conception de produits et services à l'amélioration des processus internes et à l'adoption de technologies de pointe. Cependant, notre étude se propose de canaliser cette vaste notion d'innovation pour se concentrer spécifiquement sur le domaine de la gestion de la trésorerie avec un accent particulier sur les choix de financement. En orientant notre étude vers ces dimensions insuffisamment explorées, nous visons à combler un gap identifié dans la littérature actuelle. Ainsi, l'objectif est de mettre en lumière sous un angle nouveau les mécanismes par lesquels les entreprises peuvent optimiser leurs ressources financières, se prémunir contre les risques de défaillance en saisissant les opportunités de croissance dans un environnement économique en réinvention permanente.

### 4.2. Les Modes de Financement Novateurs : Le Bouclier Indispensable pour l'optimisation de la Trésorerie et la stimulation de la Performance

#### 4.2.1. Le Capital Risque

Le capital-risque agit comme un accélérateur de performance, permettant aux entreprises de se positionner avantageusement sur leur marché, d'augmenter leur compétitivité et de réaliser une croissance soutenue. L'apport du capital-risque dans le panorama entrepreneurial ne se limite pas uniquement à une injection de fonds ; Il s'affirme un catalyseur de croissance incontournable pour les entreprises, leur permettant de surmonter les défis inhérents à leur développement.

En France, le capital-risque constitue un vecteur essentiel pour la croissance et l'innovation, ayant joué un rôle décisif dans le financement des entreprises novatrices. A ce propos, en 2015, le capital-risque français a affiché une croissance significative, avec 484 opérations représentant 1,81 milliard d'euros d'investissements, doublant ainsi le montant enregistré en 2014. Cette performance place la France au deuxième rang en Europe, juste après le Royaume-Uni, avec une part de 21 % du total (Ekeland, et al., 2016).

Dans la même veine, Yanga et ses co-auteurs ont étudié l'influence du capital-risque sur la performance des entreprises chinoises cotées au GEM (Marché des entreprises en croissance) sur la période allant de 2010 jusqu'au 2014. Les résultats empiriques indiquent que le capital-risque a une corrélation positive et significative avec la performance de l'entreprise. En revanche, les mêmes résultats démontrent que le financement par la dette peut entraver cette performance. Ainsi, les conclusions de ladite étude enseignent que l'introduction du capital-risque devrait être prioritaire dans les décisions de financement afin d'accélérer la croissance des entreprises. En outre, les entreprises doivent contrôler l'ampleur du financement par emprunt afin de ne pas augmenter le risque engendré par un endettement excessif pouvant grever les performances financières des entreprises (Yanga, et al., 2016).

Pour se prémunir contre ce risque, l'univers de financement foisonne d'alternatives novatrices qui s'imposent en tant qu'un élément catalyseur de la performance et de la croissance dont figure le capital-risque, les offres innovantes de la finance verte et des banques panafricaines, le Crowdfunding ainsi que les services offerts par la technologie révolutionnaire des blockchains.

### 4.2.2. Les Offres de Financement Innovantes Développées par les Banques Panafricaines au Service de la Performance et la Pérennité des Entreprises en Afrique

Partant de ce qui précède, les trésoriers peuvent se financer non seulement par le capital-risque mais peuvent également se servir de la variété d'options de financement innovantes offertes par le secteur bancaire au lieu de se limiter uniquement au financement bancaire classique. Dans ce contexte, le développement du secteur bancaire africain présente à son tour une profusion des modes de financement novateurs auxquels les entreprises marocaines peuvent recourir. A cet égard, le rapport économique sur l'Afrique intitulé : financements innovants pour le développement des entreprises en Afrique, NUCEA[1](2020) met en évidence que le financement innovant du secteur privé couplé à la dynamique de croissance favorise la création de valeur ajoutée pour ces entreprises. Ceci met en exergue le rôle stratégique de l'innovation dans le financement en tant qu'un levier créateur de valeur ajoutée.

Par ailleurs un financement innovant s'avère un vecteur catalyseur de la performance des entreprises africaines dans leur ensemble, et des marocaines en particulier. En outre, selon le même rapport, le système financier africain doit accélérer sa diversification en vue de créer une palette étendue d'institutions financières proposant une large gamme de produits financiers novateurs adaptés aux besoins spécifiques de l'écosystème des entreprises. Compte tenu de la prédominance des banques dans les mécanismes de financement en Afrique, celles-ci pourraient être un terrain fertile pour le développement des financements innovants afin de stimuler la croissance du secteur privé. Dans ce contexte, l'essor des banques panafricaines, dont les sièges se trouvent en Afrique du Sud, au Kenya, au Nigéria et au Maroc, a profondément transformé le secteur bancaire de diverses économies africaines par leurs produits innovants (NUCEA, 2020).

A la lumière de ce qui est avancé, l'essor des banques panafricaines et l'introduction de modes de financements innovants constituent un tournant majeur pour le paysage entrepreneurial africain, et plus particulièrement marocain. Cette évolution bancaire offre un spectre élargi d'opportunités financières, marquant le début d'une ère où les entreprises marocaines peuvent transcender les limites du financement traditionnel. En s'alignant sur cette dynamique, elles s'ouvrent à des possibilités de financement plus flexibles et adaptées à leurs besoins spécifiques, facilitant ainsi l'accès à des ressources essentielles pour leur croissance et leur compétitivité. Les offres de financement innovant, mises en avant par les banques africaines, apportent une réponse concrète aux défis de développement auxquels sont exposées les entreprises marocaines, en leur fournissant les outils nécessaires pour naviguer adroitement à travers un paysage économique de plus en plus complexe et concurrentiel. Cette synergie entre les offres financières innovantes des banques panafricaines et la stratégie de développement des entreprises marocaines s'inscrit dans une vision à long terme, où l'innovation devient un levier clé pour la création de valeur et la performance durable.

---

[1] NUCEA : Nations Unis commission économique pour l'Afrique

### 4.2.3. Les Solutions Novatrices de Financement Offertes par la Finance Verte

Dans une ère où la conscience environnementale gagne en importance, les mécanismes de financement innovants se tournent vers des solutions respectueuses de l'environnement. C'est dans ce cadre que s'ancre la notion de la finance verte, un domaine qui connaît un essor significatif et qui représente un pivot majeur pour le futur du financement durable. La Banque mondiale souligne l'urgence de cette transition : « Pour passer à une économie mondiale durable, nous devons augmenter le financement des investissements offrant des avantages environnementaux, connus sous le nom de finance verte. » (Banque mondiale, 2019 : 7). Cette affirmation met en relief l'impératif de reconfigurer les approches de financement traditionnelles pour embrasser des modèles plus verts et durables.

Opter pour la finance verte n'est plus une option mais une nécessité impérieuse pour préserver notre planète des impacts climatiques potentiellement irréversibles. Dans cette quête de durabilité, la finance se métamorphose pour inclure un éventail de produits et services financiers innovants et respectueux de l'environnement. Des instruments tels que les obligations vertes (green bonds), les droits à polluer, et les prêts hypothécaires verts (green mortgages) illustrent cette évolution, marquant une ère nouvelle d'investissements écologiques et socialement responsables. Cette transition vers une finance verte permet d'allier la rentabilité à la responsabilité en impliquant non seulement une alliance inédite entre le secteur privé et l'Etat, mais également un arbitrage judicieux entre rentabilité et risque (Fremousse et Peretti, 2021). Sous cet angle, l'intégration des pratiques novatrices de la finance verte au cœur des choix de financement ouvre la voie à une myriade d'opportunités. Elle permet non seulement d'optimiser les coûts et d'améliorer l'efficacité opérationnelle, mais aussi de renforcer la résilience des entreprises face aux pressions environnementales et sociales croissantes. Cela va au-delà de la simple optimisation financière ; c'est un investissement dans la pérennité et des entreprises et des économies africaines (Badrane et Bamousse, 2023).

Au regard des développements précédents, il devient évident que l'adoption de la finance verte représente bien plus qu'une simple tendance pour les trésoriers d'entreprises ; elle s'affirme comme un levier essentiel de leur performance financière. En tissant un lien stratégique entre les principes de durabilité environnementale et les pratiques novatrices de gestion financière, les entreprises ne se limitent pas de répondre aux attentes croissantes en matière d'éco-responsabilité ; elles bâtissent également une fondation solide pour leur propre résilience. Face aux enjeux environnementaux croissants de notre époque, cette intégration innovante offre aux entreprises une opportunité unique de se distinguer, tout en pavant la voie vers une économie globalement plus durable et respectueuse.

En somme, la finance verte ne constitue pas seulement un choix éthique mais émerge comme une stratégie financière avisée, fondamentale pour naviguer dans le paysage économique contemporain avec succès et responsabilité.

### 4.2.4. Le Financement Participatif ou le Crowdfunding

Le Crowdfunding ou le financement par la foule se distingue comme une méthode de financement révolutionnaire, offrant une alternative moderne aux voies de financement traditionnelles. Il permet de lever les fonds auprès d'un large public en vue de financer un projet (Chaaben, 2016). Initialement émergé en 2008, en réponse aux turbulences des crises économiques et financières, ce mécanisme offre trois modes de financement à savoir : le don, le prêt et l'investissement. La variété de ces modes fait du Crowdfunding un outil de financement adapté à tous types de projets (Hemdane, 2016). Le Crowdfunding représente une avancée significative dans la sphère financière ouvrant la voie vers une inclusion financière plus large et plus accessible aux différents acteurs peu servis par le financement conventionnel. Le caractère innovant du Crowdfunding en fait un outil de financement responsable, capable d'élargir considérablement le bassin de financement pour les projets peinant à se frayer un chemin à travers les canaux de financement classiques (Badrane et Bouzahir, 2023).

Il convient de souligner que le Crowdfunding a connu une croissance exponentielle à l'échelle internationale en particulier dans le contexte américain et asiatique. Néanmoins en Afrique, son développement reste encore embryonnaire. Au Maroc, le Crowdfunding évolue favorablement au profit des entreprises en particulier après l'élaboration de son cadre réglementaire à travers le Dahir n° 1-21-24 du 22 février 2021 promulguant la loi n° 15-18.

Par ailleurs, l'essor du Crowdfunding et son émergence dans les économies avancées et émergentes suggère que ce moyen de financement innovant peut devenir un outil dans les écosystèmes d'innovation de la plupart des pays. Il peut se révéler comme un pivot de la performance et de la pérennité dans un monde globalisé.

En outre, force est de constater que la majorité des travaux théoriques et empiriques ont attribué l'apport du Crowdfunding pour financer exclusivement les TPME, les start-ups et les porteurs de projets. Cependant, tout type d'entreprises y compris les grandes peuvent solliciter ce type de financement en particulier dans leur projet d'investissement qui nécessitent des fonds colossaux. A cet égard, le Crowdfunding peut se positionner comme un outil de financement innovant que les trésoriers peuvent envisager pour contourner leurs difficultés de financement notamment après la crise économique mondiale de 2008 qui a freiné l'offre des crédits bancaires et occasionné par conséquent la défaillance de nombreuses entreprises.

### 4.2.5. Les Nouvelles Frontières du Financement : Innovations et Opportunités au Cœur de la Blockchain

La digitalisation a porté ses fruits dans la sphère de financement à travers la révolution de la blockchain qui présente des offres de financement innovantes pour les entreprises afin de financer les activités d'investissement et d'exploitation. A ce propos, Alfieri et ses co-auteurs ont pointé dans leur article intitulé « la blockchain : outil innovant au service du financement des entreprises » comment cette technologie innovante peut contribuer à améliorer le financement des entreprises qu'elles soient jeunes et risquées ou matures et stables. Par le biais d'une typologie descriptive et classificatoire des utilisations de la blockchain, ils ont mis en exergue son rôle en tant qu'une technologie fondamentalement innovante qui présente de nouvelles alternatives de financements pour les entreprises en se démarquant des modes institutionnels traditionnels (Alfieri, et al., 2021).

Ainsi, cette technologie révolutionnaire fait désormais partie intégrante du paysage financier et laisse entrevoir un large champ d'opportunités de financements novateurs pour les entreprises. Par ailleurs, elle offre des avantages précieux en termes de rapidité, de transparence, de décentralisation et de réduction des coûts, ce qui peut générer des gains potentiels d'efficacité susceptibles d'optimiser la gestion de la trésorerie et d'accroitre la performance des entreprises marocaines en particulier dans un monde économique globalisé.

En définitive et sur la base de ce qui a été avancé, il apparaît clairement que les modes de financements innovants exercent un effet positif sur la performance des entreprises. Dans le même contexte, le rapport « Fonction Finance : 140 innovations au service de la croissance » exposent les différentes pratiques innovantes au service de la fonction financière et de la fonction trésorerie en particulier, dans le but d'affirmer la croissance et de booster la performance des entreprises. A ce titre, de nouvelles solutions de financement sont proposées en complément des financements bancaires classiques visant à adapter le financement aux besoins spécifiques des entreprises. Toutefois pour que ces innovations puissent apporter leurs fruits, les dirigeants doivent se familiariser avec les différents types de financements innovants et leurs caractéristiques ainsi qu'avec les différents acteurs du financement. Par ailleurs, il conviendra de faire prendre conscience que les orientations stratégiques faites par les dirigeants peuvent exercer des effets significatifs sur le besoin en fonds de roulement et la trésorerie et que cette donnée doit être prise en compte lors de l'analyse des différentes options (PWC France, 2018).

L'analyse approfondie menée dans cette étude fait ressortir la portée stratégique de la trésorerie en tant qu'une artère vitale de l'entreprise. C'est une fonction stratégique où les ressources de financement conditionnent le fonctionnement des organisations et en influent sur la performance financière, la croissance et la pérennité. C'est dans cette optique que le présent article s'est fixé pour objectif de mettre en évidence l'importance de se servir des modes de financement novateurs afin de propulser la performance tout en répondant aux défis qui se profilent dans un environnement en réinvention permanente.

Dans le panorama mondialisé actuel, ponctué d'incessantes mutations, les entreprises marocaines sont confrontées à des défis d'envergure. Les crises financières mondiales, la volatilité des taux de change, les fluctuations des prix des matières premières, les pressions inflationnistes, la concurrence exacerbée, l'accès restreint aux marchés de capitaux, ainsi que les enjeux écologiques constituent une multitude de menaces pesant lourdement sur la performance. Face à cette réalité, le recours à des stratégies de financement traditionnelles, souvent restrictives et onéreuses, montre rapidement ses limites laissant entrevoir le financement innovant comme la clef de voûte d'une navigation habile dans les tumultes de cet environnement mondialisé. L'adoption de méthodes de financement de pointe telles que le Crowdfunding, le capital-risque, les initiatives des banques panafricaines, la finance verte et les technologies blockchain, représente une véritable bouffée d'oxygène pour le tissu économique marocain.

Au-delà des défis macroéconomiques et sectoriels évoqués précédemment, une attention particulière est portée aux tribulations de trésorerie intérieures aux entreprises. Au cœur de ces préoccupations trône le financement du besoin en fonds de roulement (BFR), un paramètre critique qui influence directement l'équilibre financier des entreprises. Un BFR excessif, symptôme d'une gestion de trésorerie déséquilibrée, peut entraîner des situations de tension de liquidité pouvant mener à la cessation de paiement, voire à la défaillance.

Face à cette réalité, l'adoption de modes de financement innovants devient une stratégie indispensable pour les entreprises désireuses de sécuriser leur pérennité et de soutenir leur développement. Ces solutions financières alternatives représentent bien plus qu'une simple source de liquidités ; elles offrent la flexibilité et l'agilité nécessaire pour répondre non seulement aux fluctuations du BFR mais aussi à tout besoin de financement que les entreprises marocaines doivent combler.

En outre, dans tous les pays ayant une économie développée ou émergente, les entreprises représentent une véritable locomotive de développement. Cette assertion trouve un écho particulier au Maroc, où ces entités constituent l'armature de son économie et la valve de son sécurité sociale et économique. Néanmoins le maintien de ce rôle stratégique est tributaire d'une gestion de trésorerie innovante. A ce titre, l'introduction des pratiques novatrices à la gestion de la trésorerie est devenue un véritable impératif pour toute entreprise afin de pallier aux différents risques pouvant freiner son équilibre financier, paralyser sa croissance, inhiber sa performance financière et partant compromettre sa survie.

Par ailleurs, de par sa position unique au cœur des métiers, une fonction financière innovante a l'opportunité d'initier voire d'orchestrer la transformation des entreprises vers la performance. Dans cette même perspective, la gestion de la trésorerie en tant que son bras armé peut s'approprier des opportunités de financement novatrices permettant d'impulser la performance et renforcer sa résilience. A cet égard et avec la sophistication de la sphère financière, le foisonnement de nouvelles alternatives de financement constitue un gisement très riche pour la gestion de la trésorerie. Ainsi, les trésoriers ont accès à des solutions de plus en plus innovantes en complément des financements bancaires classiques. Ils peuvent diversifier la palette de leurs instruments de financement au lieu de se limiter uniquement au financement bancaire classique qui engendre des coûts d'endettement colossaux susceptibles de grever lourdement la performance financière.

Ainsi, dans un monde marqué par sa turbulence et ses défis constants, la fonction de la trésorerie, munie d'une panoplie de solutions de financement innovantes, endosse le rôle de boussole guidant les entreprises vers des horizons de performance accrue et de croissance consolidée. Ce virage vers une gestion financière éclairée et avant-gardiste est non seulement un impératif mais également une opportunité pour les trésoriers d'explorer et de capitaliser sur un éventail enrichi d'instruments financiers, traçant ainsi les contours d'un nouvel âge où l'innovation s'impose le maître mot d'une performance pérenne.

En définitive, l'intégration de modes de financement innovants au cœur de la fonction de trésorerie représente aujourd'hui un tournant stratégique pour les entreprises, leur ouvrant les portes vers une myriade d'opportunités. En se positionnant à l'avant-garde des pratiques financières, ces outils novateurs offrent une réponse adaptée aux défis économiques actuels, tout en privilégiant une démarche de gestion optimisée et performante. A ce titre, l'un des apports majeurs de ces instruments financiers réside dans leur capacité à réduire significativement les coûts cachés. Ces économies, alignées avec la théorie des coûts de performances cachées, transforment des dépenses latentes, telles que les manques à gagner liés à l'insuffisance de liquidité, en opportunités d'amélioration de la performance financière. Cette réduction des coûts indirects libère des ressources précieuses qui peuvent être réinvesties dans des domaines stratégiques de l'entreprise. Au-delà de la gestion efficiente des coûts, ces instruments de financement, par leur nature même, permettent aux trésoriers de déployer des stratégies plus nuancées face aux éventuels écueils financiers, notamment ceux associés au besoin en fonds de roulement, assurant ainsi la fluidité des opérations quotidiennes et la stabilité financière. En outre, ils ouvrent la voie à des investissements stratégiques qui catalysent la croissance et la compétitivité. L'engagement en faveur de la finance verte, quant à lui, permet non seulement de répondre aux enjeux écologiques mais aussi de rehausser la réputation de l'entreprise auprès des consommateurs, des investisseurs et des partenaires. Ce renforcement de l'image de marque est fondamental dans un contexte où les critères de durabilité et de responsabilité sociale pèsent de plus en plus dans les décisions d'investissement et de consommation. Enfin, la résilience et la pérennité sont des avantages intrinsèques à l'adoption de ces solutions financières. En s'adaptant aux fluctuations économiques et en anticipant les besoins futurs, les entreprises renforcent leur capacité à naviguer adroitement à travers les turbulences d'un environnement en pleine effervescence.

En résumé, les modes de financement innovants s'imposent comme des vecteurs essentiels de transformation et de progrès pour les entreprises. Ils permettent non seulement de contourner les obstacles financiers traditionnels mais aussi de tracer une voie vers une gestion de la trésorerie plus agile, plus durable et plus performante.

Dès lors, sur la base de la théorie mobilisée, les résultats obtenus et l'analyse qui en résulte, le modèle théorique élaboré est présenté comme suit :

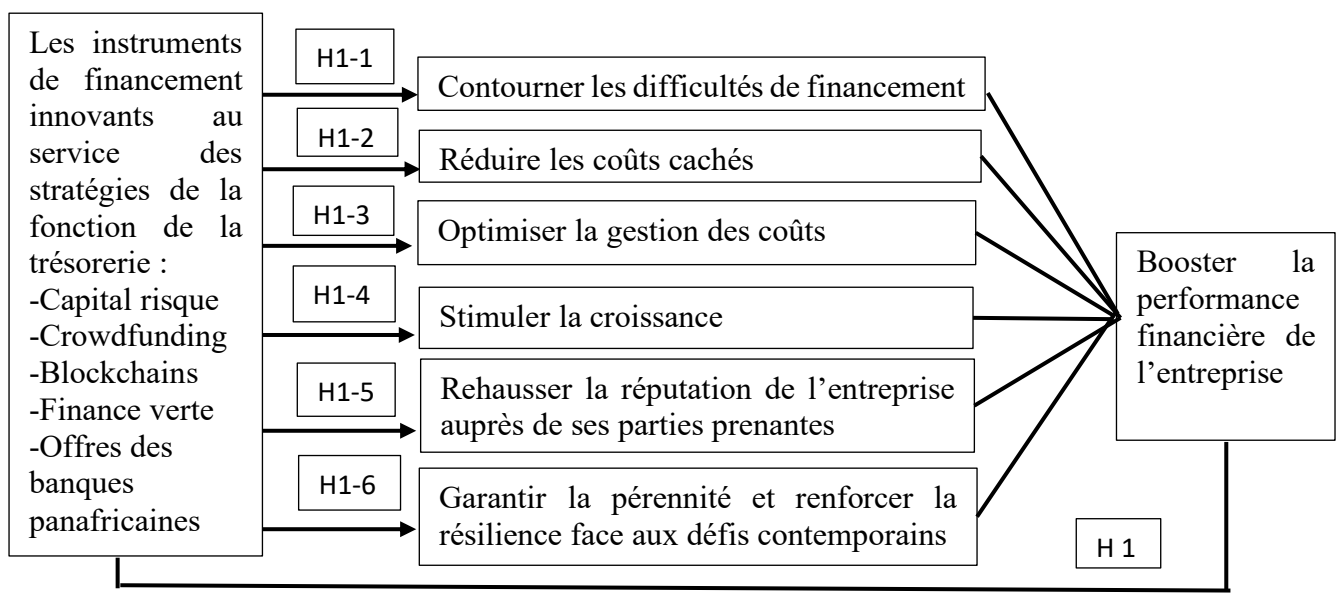

Source : les auteurs

## Conclusion

A l'issue de cette étude, le présent article avait comme objectif de répondre à notre problématique initiale en mettant en lumière la manière dont les stratégies novatrices de financement peuvent catalyser une amélioration significative de la performance financière des entreprises, les propulsant vers un avenir plus prometteur.

Loin d'être perçue comme une fonction périphérique, la gestion de la trésorerie transcende sa conception traditionnelle pour embrasser un rôle stratégique et prépondérant. Elle s'affirme le pilier sur lequel repose la dynamique de la croissance et la stabilité des entreprises.

En effet, dans un contexte économique où les crises financières ont révélé la fragilité des marchés, une gestion de trésorerie rigoureuse et avant-gardiste devient la clef de voûte de la survie et du développement des entreprises. Dans cette optique, le présent article incite les entreprises marocaines à reconsidérer leurs stratégies de financement en explorant et en intégrant des alternatives financières novatrices. Les trésoriers peuvent diversifier la palette de leurs instruments de financement au lieu de se limiter uniquement au financement bancaire classique qui engendre des coûts d'endettement colossaux susceptibles de grever lourdement la trésorerie. A cet égard, le spectre de ces innovations est large, incluant le capital-risque, le financement participatif, l'utilisation stratégique des blockchains ainsi que l'exploration de services financiers offerts par la finance verte et les banques panafricaines innovantes. Chacune de ces avenues offre aux entreprises des opportunités uniques pour renforcer leur résilience, leur offrant les moyens de naviguer avec agilité dans des environnements imprévisibles.

Les contributions de cette étude s'inscrivent dans une triple perspective académique, managériale et socioéconomique. Dans le champ académique, les résultats théoriques et les travaux empiriques traitant l'impact d'un mode de financement innovant sur la performance financière des entreprises restent limités, particulièrement dans le panorama marocain. Ainsi, le présent article avait pour ambition de creuser au maximum pour pouvoir trouver et expliquer ce lien susceptible d'exister en vue d'enrichir la réflexion autour d'un domaine disciplinaire encore peu investigué. Parallèlement, sur le plan managérial, cette étude permet aux directeurs financiers et aux trésoriers d'entreprises de s'affranchir des pratiques conventionnelles de gestion de la trésorerie afin qu'ils puissent comprendre le rôle décisif de l'innovation dans le renforcement de la résilience de leurs firmes. D'où la nécessité de déployer des pratiques innovantes en termes de financement en vue de triompher sur les aléas d'un environnement globalisé.

Par ailleurs, sous une perspective socio-économique, l'objectif du présent article est d'atteindre les objectifs de développement durable visant à favoriser une croissance inclusive. En effet, les entreprises forment le noyau vital de l'économie nationale. À ce titre, il est déterminant pour les trésoriers de garantir des financements innovants susceptibles de maintenir la pérennité de leur firme et par extension soutenir la création d'emplois. Ce qui permet d'accélérer la croissance socio-économique en renforçant le rôle du Maroc en tant que locomotive de développement continental.

Compte tenu de ce qui précède, les apports de cette étude s'annoncent prometteuses. Néanmoins, la confirmation des résultats théoriques doit être nuancés par une étude exploratoire. En effet, comme tout effort de recherche, ce travail n'est évidemment pas dénué de certaines limites, toutefois ces limites élargissent nos horizons de recherches sur d'autres pistes à explorer. A ce titre, une étude comparative à visée exploratoire réunissant des entreprises performantes de différents pays peut être menée afin d'identifier les meilleures pratiques innovantes de financement qui sont susceptibles de stimuler la performance afin de les préconiser au contexte marocain.  En outre, ce travail ouvre également de nouvelles voies de recherches quant au rôle des gouvernements africains dans l'instauration d'un écosystème sain pour tout type d'entreprises afin de tirer pleinement profit du potentiel des financements innovants. Ceci témoigne que la performance des entreprises et le développement inclusif peuvent aller de pair pour façonner un avenir prometteur et durable. Dans le même ordre d'idées, d'autres études peuvent être envisagées pour mettre en avant que l'innovation intervient non seulement dans le choix du mode de financement, mais s'étend pour couvrir également l'établissement en amont des prévisions de trésorerie et des besoins de financement fiables susceptibles d'orienter judicieusement les décisions stratégiques des trésoriers vers des modes de financements innovants et performants. Dans ce contexte, des études futures pourraient explorer en profondeur comment l'intégration de la technologie révolutionnaire de l'intelligence artificielle transforme les stratégies de trésorerie, en facilitant une sélection plus éclairée et stratégique des modes de financement.

## BIBLIOGRAPHIE